\documentclass{article}

\usepackage{arxiv}

\usepackage[utf8]{inputenc} 
\usepackage[T1]{fontenc}    
\usepackage{hyperref}       
\usepackage{url}            
\usepackage{booktabs}       
\usepackage{amsfonts}       
\usepackage{nicefrac}       
\usepackage{microtype}      
\usepackage{lipsum}
\usepackage{graphicx}
\usepackage{amsmath}
\usepackage{algorithm}
\usepackage{algpseudocode}
\usepackage[running]{lineno}
\usepackage{multirow}
\usepackage{acronym}
\usepackage{tabularx}
\usepackage{orcidlink}
\graphicspath{ {./images/} }

\title{Labeling NIDS Rules with MITRE ATT\&CK Techniques: Machine Learning vs. Large Language Models}

\author{
 Nir Daniel \orcidlink{0000-0001-5310-2324} \\
  Ben-Gurion University of the Negev\\
  Cyber@BGU\\
  Israel\\
  \texttt{nirdanie@post.bgu.ac.il} \\
   \And
 Florian Klaus Kaiser \\
  Agnostic Intelligence AG\\
  Switzerland\\
  \texttt{kaiserf@agnostic-intelligence.com} \\
  \And
 Shay Giladi, Sapir Sharabi, Raz Moyal, Shalev Shpolyansky \\
  Ben-Gurion University of the Negev\\
  Israel\\
  \texttt{\{shaygila,sapsh,razmoy,maisha\}@post.bgu.ac.il} \\
  \And
 Andres Murillo \\
  Fujitsu\\
  Japan\\
  \texttt{andres\_murillo@sutd.edu.sg} \\
  \And
 Aviad Elyashar \orcidlink{0000-0002-0918-0146} \\
  Shamoon College of Engineering\\
  Cyber@BGU\\
  Israel\\
  \texttt{aviade@post.bgu.ac.il} \\
  \And
 Rami Puzis \orcidlink{0000-0002-7229-3899} \\
  Ben-Gurion University of the Negev\\
  Cyber@BGU\\
  Israel\\
  \texttt{puzis@bgu.ac.il} \\
}

\begin{document}
\maketitle
\begin{abstract}
Analysts in Security Operations Centers (SOCs) are often occupied with time-consuming investigations of alerts from Network Intrusion Detection Systems (NIDS).
Many NIDS rules lack clear explanations and associations with attack techniques, complicating the alert triage and the generation of attack hypotheses. 
Large Language Models (LLMs) may be a promising technology to reduce the alert explainability gap by associating rules with attack techniques. 
In this paper, we investigate the ability of three prominent LLMs (ChatGPT, Claude, and Gemini) to reason about NIDS rules while labeling them with MITRE ATT\&CK tactics and techniques.  
We discuss prompt design and present experiments performed with 973 Snort rules. 
Our results indicate that while LLMs provide explainable, scalable, and efficient initial mappings, traditional Machine Learning (ML) models consistently outperform them in accuracy, achieving higher precision, recall, and F1-scores. 
These results highlight the potential for hybrid LLM-ML approaches to enhance SOC operations and better address the evolving threat landscape.
\end{abstract}


\section{Introduction}
\ac{NIDS}, such as Snort~\footnote{\url{https://www.snort.org/}} or Suricata~\footnote{\url{https://suricata.io/}}, analyze network traffic to identify and mitigate potential threats.
In this way, \ac{NIDS} rules provide an efficient means of identifying specific attack procedures~\cite{chakrabarti2010study}.  
NIDS-generated alerts must be interpreted by security analysts to extract relevant information to understand ongoing attacks. 
To mitigate detected attacks, a security analyst needs to construct valid hypotheses regarding attack techniques and attackers' intentions, generating actionable insights~\cite{elitzur2019attack}.
Leveraging on such actionable insights, security analysts are empowered to take appropriate countermeasures.

The increasing sophistication and diversity of cyber attacks necessitates a complex rule-based approach in \ac{NIDS} (i.e. leading to a significant number of rules \ac{NIDS} operate on).
The plethora of different \ac{NIDS} rules analysts need to understand and the lack of links between those rules (e.g. Snort) and comprehensive Cyber Threat Intelligence (CTI) requires significant skills, attention, and cyber security expertise~\cite{lin2022attack}.
Given the scarcity of experienced security analysts, the need for labeled \ac{NIDS} rules supporting (and partially automating) the process of identifying attacks, detecting adversary behavior, and hypothesize about probable next steps is highlighted in recent publications~\cite{bagui2023introducing}.

To reduce human effort and help analysts keep up with malicious actors, companies are increasingly incorporating \ac{AI} into their security workflows~\cite{Sentonas_2023}. 
A primary application of \ac{AI} is automating the investigation of security events, which helps alleviate alert fatigue among analysts. 
Beyond this, \ac{AI} presents numerous opportunities to enhance defensive security operations. 
Hereby, \ac{LLMs} are being used to automate various cyber security tasks that previously relied on human effort~\cite{tornberg2023chatgpt}~\cite{long2023evaluating}.
By automating these processes, \ac{AI} allows security professionals to focus on higher-level tasks, adding significant value to organizations and society.


In this paper, we utilize \ac{AI} to enrich \ac{NIDS} rules with relevant MITRE ATT\&CK Techniques\footnote{\url{https://attack.mitre.org/}}, easing the analysts work in extracting actionable insights into ongoing attacks and craft hypotheses.  
 
We investigate the feasibility of using three open-accessible \ac{LLMs} to assist cyber security professionals.
The research problem targeted within this work is formulated by Guerra et al.~\cite{guerra2022datasets} and Gjerstad et al.~\cite{gjerstad2022generating}, demanding for novel methods for labeling high volumes of network traffic related data with high quality and speed.
Namely, we explore, optimize and evaluate the usage of ChatGPT~\footnote{\url{https://chatgpt.com/}}, Claude~\footnote{\url{https://claude.ai/}}, and Gemini~\footnote{\url{https://gemini.google.com/}} to automate the labeling of Snort rules with MITRE ATT\&CK tactics and techniques, thereby suggesting information about the current state of a cyber attack.

Our main contributions are as follows:
\begin{itemize}
    \item A dataset of 973 labeled Snort community \ac{NIDS} rules;
    \item formalizing \ac{NIDS} rule labeling as a conditional text generation problem, treating it as a maximization task;
    \item introducing a workflow for employing \ac{LLMs} to labeling \ac{NIDS} rules with MITRE ATT\&CK tactics and techniques and generating an automated \ac{ML}-based labeling procedure;
    \item three \ac{ML} models for labeling \ac{NIDS} rules, automatically generated and trained by \ac{LLMs}.
\end{itemize}

Using \ac{LLMs} for labeling \ac{NIDS} rules provides two main advantages: 
Firstly, it is possible to explain each technique suggestion. 
The provided explanations and reasoning given through the \ac{LLMs} can be especially beneficial for analysts with limited expertise in cyber security (e.g., limited experience).
Second, when using \ac{LLMs} we utilize the knowledge base on which they are trained, which contains cyber security knowledge from a wide range of different data sources including incident reports but also hacker chatter, etc.
\ac{ML}-models on the other side frequently provide superior performance within the tasks provided however, while results are frequently limited with respect to their explainability.


The rest of the paper is structured as follows. Section~\ref{sec:back} gives relevant background. Section~\ref{sec:related_work} presents related work. In section~\ref{sec:meth} the methods, dataset and experiments are presented. In section~\ref{sec:res} we introduce the results. Section ~\ref{sec:disc} provides the  discussion. Section~\ref{sec:conc} contains a summary of the main insights and provides a research outlook.

\section{Background}
\label{sec:back}
\subsection{Cyber Threat Intelligence}
\label{sec:back_cti}
\ac{CTI} is a proactive approach in computer and network security~\cite{palacin2021practical}, focusing on the collection and analysis of data to derive actionable insights into potential or ongoing attacks. 
These insights improve decision-making processes by enabling the selection of appropriate defensive measures~\cite{chismon2015threat}. 
As outlined by Chismon et al.~\cite{chismon2015threat}, \ac{CTI} can be categorized into four types based on focus and depth:

\begin{itemize}
    \item \emph{Strategic \ac{CTI}} Aimed at non-technical audiences~\cite{haddad2023automated}, this high-level intelligence assists organizational leaders in understanding the broader implications of cyber activities, including potential risks and their impacts.
    \item \emph{Operational \ac{CTI}} Provides detailed insights on impending attacks, consumed primarily by senior security staff, such as heads of incident response teams. This intelligence supports day-to-day decision-making, but access is often limited to advanced organizations or nation-states.
    \item \emph{Tactical \ac{CTI}} Known as \ac{TTPs}, this intelligence details adversary methodologies and is used by \ac{SOCs} to test and enhance defensive measures
    \item \emph{Technical \ac{CTI}} Provides raw data, such as IP addresses or hash values, which are time-sensitive and must be used promptly as they can quickly become obsolete 
\end{itemize}

Another categorization of \ac{CTI} is based on its source, which can be either \textbf{network-based} or \textbf{host-based}.

Low-level \ac{CTI}, including \ac{IoC}, is especially valuable for automating cyber security processes, enabling more efficient decision-making~\cite{daszczyszak2019ttp}.
Tools such as \ac{TH}, \ac{EDR}, and \ac{IDS} often rely on technical and tactical \ac{CTI} to identify and address potential threats~\cite{kaiser2023attack}.
A great potential is hereby based on the ability to expand the levels of automation~\cite{kaiser2023attack} and aid organizations gaining visibility of the threat landscape, identify attacks and the \ac{TTPs} confronted with, as well as to effectively respond to an attack~\cite{liao2016acing}.

\label{sec:back-mitre_attack}
The MITRE ATT\&CK framework is an essential resource for \ac{CTI}, providing a curated knowledge base of adversarial behavior, which includes detailed mappings of post-compromise tactics and techniques across various platforms~\cite{strom2020mitre}.
According to Strom et al.~\cite{strom2020mitre}, the ATT\&CK framework centers on understanding how external adversaries infiltrate and act within computer networks. 
It therefore serves as an extensive database of tactical \ac{CTI}, encompassing post-compromise adversary \ac{TTPs} across various operating systems, including Windows, Linux, and macOS.
Additionally, it spans multiple technological domains such as enterprise environments, mobile devices, cloud systems, and \ac{ICS}. 
As of November 2024, the ATT\&CK framework includes 203 attack techniques and 453 sub-techniques associated with 14 tactics for enterprise systems and 84 techniques linked to 12 tactics specific to \ac{ICS}.

\subsection{Intrusion Detection Systems}

IDS can be differentiated to \ac{HIDS} and \ac{NIDS}~\cite{landauer2022framework}.
The visibility coverage of different \ac{IDS} depends especially on the data sources they analyze.
NIDS, for example, offer only limited visibility inside the host machine~\cite{othman2018survey}.
Therefore, only a limited set of attack techniques is detectable with high confidence relying on \ac{NIDS}.
Likewise, \ac{HIDS} provide restricted visibility by being limited on the analysis of system logs.
In an effort to systemize the analysis of data log source quality, visibility coverage, and detection coverage, Detect Tactics, Techniques \& Combat Threats ($DeTT\&CK$) was developed\footnote{\url{https://github.com/rabobank-cdc/DeTTECT}}. 
$DeTT\&CK$ provides a useful means to analyze the visibility coverage of different intrusion detection systems and techniques identifiable through investigations of specific data sources.

\begin{figure}[!h]
\includegraphics[width=\textwidth]{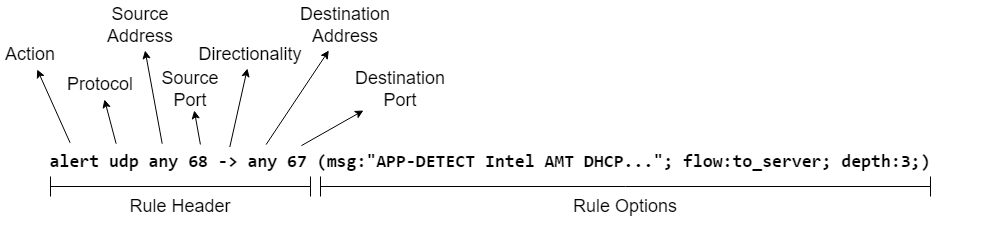}
\caption{Snort rule example} \label{fig1}
\end{figure}

Snort\footnote{\url{https://www.snort.org/}} is a lightweight \ac{NIDS} built on the Libpcap library, offering efficient packet filtering capabilities~\cite{peng2012design}. 
Snort rules, which are straightforward and easily interpreted, are divided into two primary components~\cite{khamphakdee2014improving} (see Fig.~\ref{fig1}): The first component, known as the header, specifies attributes such as action, protocol, source and destination addresses, source and destination ports, and traffic direction. 
The second component, known as the rule options, contains a list of keyword and argument pairs enclosed within parentheses. 
These elements collectively enable Snort to detect and respond to specific network-based attack patterns effectively.

\unskip

\subsection{Large Language Models}

\ac{GPTs} are among the most advanced \ac{LLMs} currently available, known for their ability to process and analyze complex tasks across various fields. 
This study leverages ChatGPT-4 
along with other state-of-the-art \ac{LLMs}, including Claude 
and Gemini 
, to enhance cyber security workflows. 
These models, with their advanced reasoning capabilities, have demonstrated potential in automating expert-driven processes~\cite{tornberg2023chatgpt}.

In cyber security, \ac{LLMs} are gaining recognition for their dual utility in both defensive and offensive applications~\cite{motlagh2024large}. 
On the offensive side, \ac{LLMs} have raised concerns due to their ability to facilitate malicious activities, such as generating phishing emails or creating sophisticated attack scripts~\cite{gabriela2023chatgpt}. 
On the defensive side, however, these models have emerged as powerful tools for automating labor-intensive tasks~\cite{liu2024logprompt}. 
Their vast training datasets, which include security-related knowledge, allow them to support cyber security experts by generating actionable insights and assisting in complex decision-making processes.

This work explores a further use case in analyzing the feasibility of using \ac{LLMs} to label \ac{NIDS} rules with MITRE ATT\&CK techniques and furthermore provide contextual explanations for each label. 
By leveraging the comprehensive training and reasoning capabilities of models such as ChatGPT, Claude, and Gemini, we aim to evaluate and compare their effectiveness in enriching \ac{CTI} workflows.
This marks a significant advancement in the adoption of \ac{LLMs} as a scalable and efficient means to assist security analysts in mitigating and understanding ongoing threats.

\section{Related Work}
\label{sec:related_work}
Ishibashi et al.~\cite{ishibashi2022generating} highlights the need to implement \ac{AI}-powered \ac{NIDS} and construct labeled training datasets. 
Gjerstad~\cite{gjerstad2022generating} highlights the lack and motivates the generation of labeled cyber security data sets (i.a., \ac{CTI} labeled reports, and labeled host-based, network-based data). 

The increasing capabilities of \ac{LLMs} lead to rising adoption in different fields.
Due to the high volume and speed of \ac{CTI} sharing, automation of labeling can be considered a necessary means for staying on top of potential perpetrators of cyber attacks.


\subsection{Labeling reports with CTI}

Husari et al.~\cite{husari2017ttpdrill} present TTPDrill, a method for extracting \ac{TTPs} from unstructured text.
As part of the extraction process, TTPDrill uses a text-mining approach that combines natural language processing and information retrieval.
Another tool for extracting \ac{TTPs} from reports named rcATT is introduced by Legoy et al.~\cite{legoy2020automated}.
This tool is based on \ac{MCML-C}, trained on reports mapped to \ac{TTPs}.
Mendsaikhan et al.~\cite{mendsaikhan2020automatic} adopt \ac{ML-C} on vectors from vulnerability descriptions (reports) and linking those vectors to adversary techniques.
Extractor~\cite{satvat2021extractor} furthermore is capable of extracting attack behaviors as provenance graphs from unstructured text.
TIM~\cite{you2022tim}, represents a threat context-enhanced \ac{TTPs} intelligence mining framework operating on unstructured data. 
With AttacKG Li et al.~\cite{li2022attackg} introduced a similar tool that extracts techniques from reports. 
The most recent work on the extraction of \ac{TTPs} from reports is TTPHunter~\cite{rani2023ttphunter} which targets APT reports taking advantage of SecureBERT~\cite{aghaei2022securebert}. 

The above methods are intended for extracting \ac{TTPs} from unstructured text. 
However, it is unclear how they will perform on structured \ac{CTI} sources and on input that demands for a high level of expertise and context information such as \ac{NIDS} rules (e.g. an understanding of benign and malicious network flows and link techniques that generate such network traffic).

\subsection{Labeling of host-related data with CTI}

Landauer et al.~\cite{landauer2022framework} introduce a method targeting the labeling within \ac{HIDS}.
The method proposed aims at labeling system logs.

Gabrys et al.~\cite{gabrys2024using} integrate \ac{LLMs} and \ac{ML} to map host-based Wazuh \ac{IDS} rules to MITRE ATT\&CK techniques, achieving high classification accuracy and enhancing alert interpretability. 
Their focus on host-based rules, which describe processes and endpoint events, aligns more directly with ATT\&CK techniques compared to network-based rules. 

Compared to those works presented, our work focuses on network-based intrusion detection rather than on host-based intrusion detection. 
Although the work presented by Gabrys et al.~\cite{gabrys2024using} is comparable with respect to the use of \ac{LLMs} and \ac{ML} the entirely different focus on \ac{NIDS} rules differentiates the works substantially.

\subsection{Labeling of network-related data with CTI}
\subsubsection{Labeling Network Packets}
According to McPhee~\cite{Mcphee2020MethodsTE}, approaches for network-based detection of ATT\&CK techniques based on network sensors, are focused only on Windows-specific protocols, and do not cover different types of systems (e.g., \ac{ICS}). 
For this purpose, McPhee~\cite{Mcphee2020MethodsTE} demonstrated the use of the network monitoring system Zeek\footnote{\url{https://zeek.org/}} for the detection of techniques in those environments. 
While this can be very useful, his approach involved manually defining the detection method for each individual technique, which requires expert knowledge and represents a time-consuming task.
Arafune et al.~\cite{arafune2022design} specifically target \ac{ICS} and develop an automated \ac{TH}, which detects attacks in network traffic using open-source tools. 
In their approach, they are linking network traffic to attack techniques, however, their method is signature-based, meaning that a human analyst still needs to perform the work of labeling at least one time for each signature, in order to be able to identify the techniques matching that signature.

Garcia, Valeros~\cite{garcia2023towards} introduce a tool to label Zeek network flows with parts of MITRE ATT\&CK framework.
Also, Masumi et al.~\cite{masumi2021towards} focus on labeling network packets to attack information.
Gjerstad~\cite{gjerstad2022generating}  presents an approach for labeling network data sets using MITRE CALDERA, a tool developed for experts to test the security of their systems. 
The approach is based on matching techniques already labeled in the CALDERA reports to attack traffic generated by simulations.
Furthermore, Bagui et al.~\cite{bagui2023introducing} introduce a dataset containing network traffic which is labeled using the MITRE ATT\&CK framework. 
Within the work, a network data set is crafted using Zeek and PCAPs. 
The mapping is based on a rule-based mapping process where link to a specific ATT\&CK technique are generated using pre-configured mappings (i.e. already labeled mission logs) and does not describe the process of labeling it. 
A further tool called RADAR~\cite{sharma2022radar} can identify malicious behavior in network traffic. 
As part of its methodology, it detects \ac{TTPs} from network traffic using both feature-based and heuristic-based detection rules. 
One of RADAR's limitations is that the feature-based rules are generated manually for every technique the system needs to support, which results in a small number of techniques currently supported by the system - only 17 techniques.

Jüttner et al.~\cite{juttner2024chatids} propose ChatIDS, a system that employs \ac{LLMs} to translate \ac{IDS} alerts into intuitive explanations and actionable countermeasures for non-expert users, aiming to improve cyber security in home and small network environments. 
By focusing on accessibility, ChatIDS simplifies alerts to help users understand threats and respond appropriately without requiring technical expertise.
In contrast, our work uses \ac{LLMs} to label \ac{NIDS} rules with MITRE ATT\&CK techniques, providing detailed, actionable insights for professional analysts. 
While ChatIDS emphasizes ease of use and comprehensibility, our solution is designed to enhance the efficiency and precision of high-level \ac{CTI} workflows in complex, large-scale cyber security operations (labeling with standardized machine readable labels and explainable mappings).
The evaluation methods further highlight these differences. 
ChatIDS relies on qualitative feedback from interdisciplinary experts to assess usability and practical applicability for non-technical users. 
In contrast, our study uses quantitative metrics—precision, recall, and F1-score—to evaluate the accuracy and scalability of \ac{LLMs}, reflecting its focus on meeting the demands of expert-driven, operational environments.

\subsubsection{Network Intrusion Detection Rules Labeling}

The work closest to ours is presented by Lin et al.~\cite{lin2022attack}.
They propose a mechanism for labeling \ac{NIDS} rules with 12 attack tactics using text mining and machine learning (esp. \ac{DT}, \ac{KNN}, \ac{SVM}, and \ac{RF}), in order to aid experts during the \ac{TH} process.
They reach a F1-score of approximately 0.9.
In their work, labeling of rules is restricted to tactics significantly limiting its usability.
Furthermore, it is easier than labeling rules with techniques, since currently, there are only 26 tactics, whereas there are at least 203 techniques (see section~\ref{sec:back_cti}).
Furthermore, Daniel et al.~\cite{daniel2023labeling}, provide a proof of concept demonstrating the potential of ChatGPT as a tool for enriching \ac{NIDS} rules with MITRE ATT\&CK techniques.

We expand the approach presented in Daniel et al.~\cite{daniel2023labeling} by (1) expanding the dataset of labeled rules, and provide it as a benchmarking, training or comparing tool for other researches in the field, (2) incorporating additional \ac{LLMs}, specifically Claude and Gemini, to evaluate their effectiveness in performing similar tasks, (3) expanding labeling workflow to cover \ac{NIDS} rules labeling comprehensively relying on MITRE ATT\&CK (labeling \ac{NIDS} rules with tactics and techniques). 
Furthermore, we formalize the task of labeling \ac{NIDS} rules as a conditional text generation problem and contribute a comprehensive prompt engineering approach.

Besides investigating the ability of \ac{LLMs} to label \ac{NIDS} rules, we introduce three \ac{ML}-models for \ac{NIDS} rule labeling. 
This broader investigation enables a comparative analysis of \ac{LLMs} and \ac{ML}-models capabilities and provides deeper insights into the applicability of \ac{AI} for automating cyber security processes.

Table~\ref{tab:relwork} provides an overview of related work highlighting some key extensions we incorporated within our work~\footnote{For increasing readability we use the following abbreviations:\\ \ac{AT} \\ \ac{E} \\ \ac{LCA} \\ \ac{NER} \\ \ac{TA} 
 \\ \ac{T} }.

\begin{table}[H]
\caption{Related work \label{tab:relwork}}
\newcolumntype{C}{>{\centering\footnotesize\arraybackslash}X}
\newcolumntype{P}[1]{>{\raggedright\footnotesize\arraybackslash}p{#1}}
\newcolumntype{Z}[1]{>{\centering\footnotesize\arraybackslash}p{#1}}
\begin{tabularx}{\textwidth}{P{2.9cm}Z{1.5cm}CZ{1.5cm}CZ{1.6cm}Z{1.5cm}}
\toprule
\textbf{Citation} &  & \textbf{Labeling}  & & & \textbf{Methodology} &\\
&Input & Output & Evaluation & LLM & ML & Other\\
\midrule
\hline

Husari et al.~\cite{husari2017ttpdrill} & Reports & \acs{TA}; \acs{T} & Quantitative & - & \acs{SVM} & \acs{TF-IDF}\\
Legoy et al.~\cite{legoy2020automated} & Reports & \acs{TA}; \acs{T} & Quantitative & - & \acs{SVM}; \acs{ML-C}; \acs{MCML-C} &\acs{TF}\label{acro:TF};\acs{TF-IDF}; Word2Vec \\
Mendsaikhan et al.~\cite{mendsaikhan2020automatic} & Reports & \acs{T} & Quantitative & - & \acs{ML-C} & \acs{TF-IDF}; Bag of Words \\
Satvat et al.\cite{satvat2021extractor} & Reports & \acs{T} & Quantitative & BERT & - & - \\
You et al.~\cite{you2022tim} & Reports & \acs{TA}; \acs{T} & Quantitative & - & - & \acs{TF-IDF} \\
Rani et al.~\cite{rani2023ttphunter} & Reports & \acs{TA}; \acs{T} & Quantitative & SecureBERT & - & - \\
Li et al.~\cite{li2022attackg} & Reports & \acs{T} & Quantitative & - & - & Regex; NER; LCA\\
Landauer et al.~\cite{landauer2022framework} & Host-based logs & \acs{TA}; \acs{T} & Qualitative & - & - & Rule-based\\
Gabrys et al.~\cite{gabrys2024using} & HIDS-rules & \acs{TA}; \acs{T} & Quantitative & BERT & - & - \\
McPhee~\cite{Mcphee2020MethodsTE} & Network traffic (Zeek) & \acs{T} & Quantitative & - & - & Rule-based\\
Arafune et al.~\cite{arafune2022design} & Network traffic & \acs{TA}; \acs{T} & Quantitative & - & \acs{SVM} & - \\
Garcia, Valeros~\cite{garcia2023towards} & Network traffic & \acs{T} & Qualitative; Quantitative & - & - & Rule-based\\
Masumi et al.~\cite{masumi2021towards} & Network traffic & \acs{AT}& Quantitative & - & - & Correlation analysis\\
Gjerstad~\cite{gjerstad2022generating} & Network traffic & \acs{T} & Qualitative & - & - & Rule-based\\
Bagui et al.~\cite{bagui2023introducing} & Network traffic & \acs{TA}; \acs{T} & Not specified & - & - & Rule-based\\
Sharma et al.~\cite{sharma2022radar} & Network traffic & \acs{TA}; \acs{T} & Quantitative & - & - & Rule-based \\
Jüttner et al.~\cite{juttner2024chatids} & IDS alerts (Surricata, Snort, Zeek) & \acs{E} & Qualitative & ChatGPT & - & -\\
Lin et al.~\cite{lin2022attack} & NIDS-rules & \acs{TA} & Quantitative & - & \acs{SVM}; \acs{RF}; \ac{DT}; \ac{KNN} & \acs{TF-IDF} \\
Daniel et al.~\cite{daniel2023labeling} & NIDS-rules & \acs{T} & Quantitative & ChatGPT & - & Keyword-based\\
Our work & NIDS-rules & \acs{TA}; \acs{T} & Quantitative & ChatGPT; Claude; Gemini & \acs{SVM}, \acs{RF}, \acs{GBM} & \acs{TF-IDF} \\
\hline
\bottomrule
\end{tabularx}
\end{table}

\section{Materials and Methods}
\label{sec:meth}
\begin{figure}[!h]
\includegraphics[width=\textwidth]{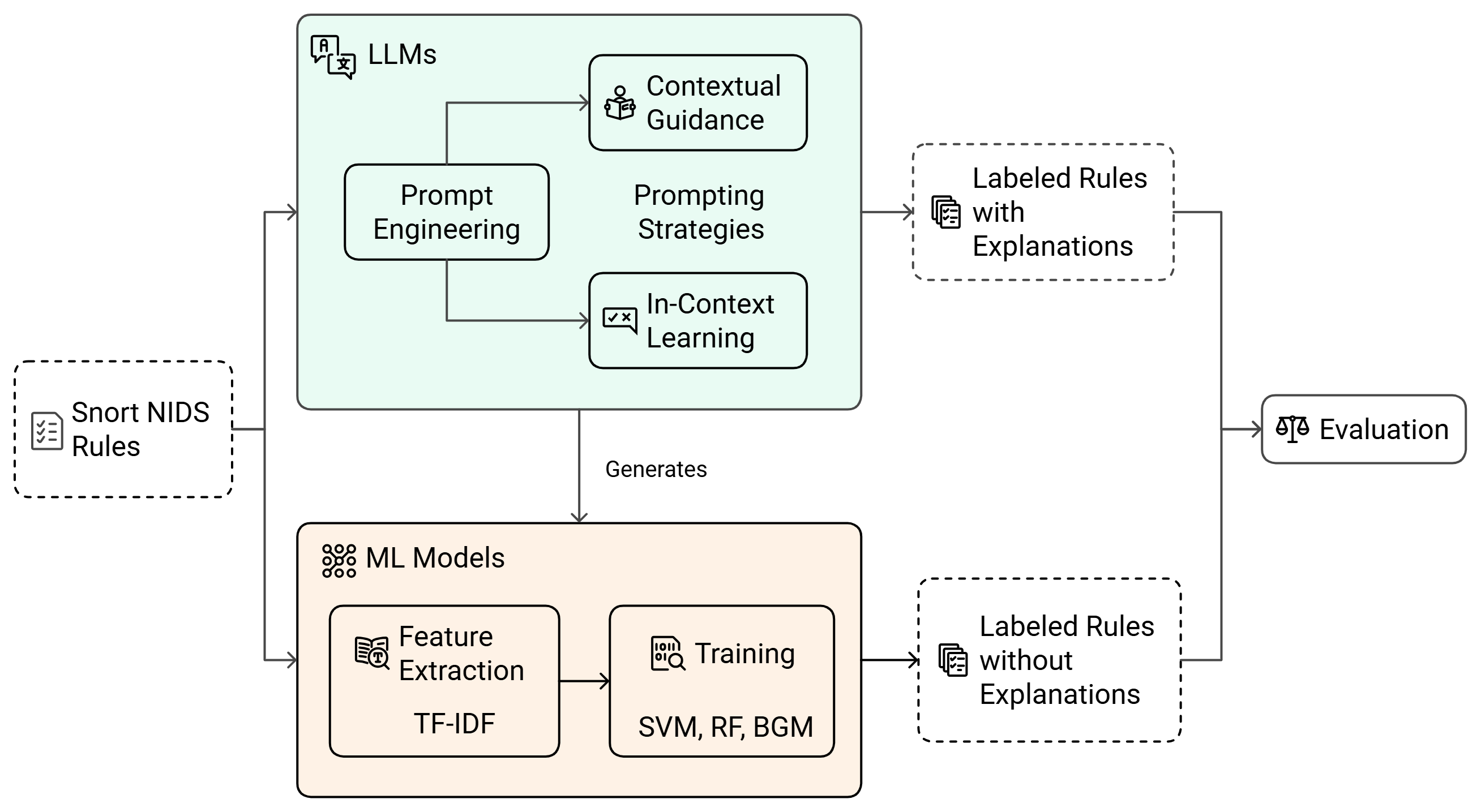}
\caption{Materials and methods overview} \label{fig2}
\end{figure}

Figure~\ref{fig2} presents an overview of the stages discussed in this section.
We present two approaches for labeling \ac{NIDS} rules with MITRE ATT\&CK techniques. 
In the first approach we use \ac{LLMs} and test different combinations of prompting strategies as explained in section~\ref{sec:prompts}. 
In the second approach (see section~\ref{sec:ml}), we test \ac{ML} models generated by the \ac{LLMs} themselves. 
Based on the technique mapping, we label the \ac{NIDS} rule with the relevant ATT\&CK tactic(s) associated with the ATT\&CK technique(s).
While the aim of this work is to label \ac{NIDS} rules with ATT\&CK techniques, this procedure allows to compare with related work.
The rest of this section covers the dataset collection, experiments and evaluation metrics used to compare the performance of the two approaches.

\unskip

\subsection{Prompt Engineering for the NIDS Labeling Task}
\label{sec:prompts}

\subsubsection{Problem Definition}

In the realm of using \ac{LLMs}, prompt engineering is a critical task that involves crafting effective cues to guide \ac{LLMs} in generating desired text outputs. 
Given the probabilistic nature of outputs generated by \ac{LLMs} a fundamental challenge in prompt engineering is understanding the probability of a language model generating a specific text given a particular prompt. 
This probability can be formalized as a conditional probability.
The \ac{LLM} based labeling can hence be regarded as a conditional text generation problem.

Let $p$ be a prompt.
The goal of engineering a prompt for \ac{LLM}-based labeling in the context of labeling \ac{NIDS} rules, is to create a well-defined and efficient prompt $p$ (cue or input given to the LLM) from an input space $P$ that enables accurate labeling of \ac{NIDS} rules (output).
With respect to this, the output of the \ac{LLM} can be modeled as a function $f(p\in P,x\in X)$
where $x\in X$ is the \ac{NIDS} rule to be labeled from a set of \ac{NIDS} rules $X$.
The problem is formulated as a maximization problem as follows.

\begin{linenomath}
\begin{equation}
    max_{X,V} g(f(p,x),v)
\end{equation}
\end{linenomath}

where $g$ describes the applicable evaluation metric (e.g., F1-score for the labeling task) and $V$ is the ground truth.
$g$ is thereby defined as the labeling score between the predicted label $v^{h}_i\in V$ and the ground truth $v_i\in V$ for a given input $x_i$ and prompt $p$.

The evaluation is based on the dataset $D$, where $x,v\in D$ is an instance of the dataset with the label $x$ and the true label $v$
where the evaluation metric is calculated based on the average labeling score.

\begin{linenomath}
\begin{equation}
    \frac{1}{|D|} \sum_{x,v\in D}g(f(p,x),v)
\end{equation}
\end{linenomath}

\subsubsection{Prompt Template Generation}
The function describing the output of the LLM can be specified as a function $Pr_{LLM}(y|p)$ where the probability $Pr$ of generating a text $y$ is modeled as a conditional probability depending on $p$, where $p$ can be decomposed to its key components.
The components of $p$ are thereby represented by the task specification $S$, contextual information $C$, and methodological guidance $M$. 

\begin{linenomath}
\begin{equation}
    Pr_{LLM}(y|p) = Pr_{LLM}(y|S,C,M)
\end{equation}
\end{linenomath}

Each component is optimized based on prompt engineering techniques.
The basic task specification $S$ defines the desired output content.
Direct questioning was employed to ask explicitly which MITRE ATT\&CK technique(s) correspond to a given \ac{NIDS} rule. 
We use the meta prompting technique of self-refinement prompting~\cite{madaan2024self} to optimize the basic task specification.
In this way, the prompts were refined iteratively based on initial model responses to enhance clarity and reduce ambiguity.

Additionally, contextual guidance ($C$) was provided within the prompts to direct the model's focus toward relevant techniques to the \ac{NIDS} rules.
The techniques list (T) provided to the \ac{LLM} is the complete list of techniques included in MITRE ATT\&CK.


With regard to the methodological guidance ($M$) we tested the use of competition questioning and few-shot \ac{ICL}.
Algorithm~\ref{alg:competition_questioning} describes competition questioning - this method involves dividing list of MITRE ATT\&CK techniques into smaller, more manageable batches and conducting multiple rounds of questioning to refine the models' labeling choices.
The complete set of MITRE ATT\&CK techniques was divided into 11 batches, and each \ac{LLM} was tasked with labeling a given \ac{NIDS} rule using only the techniques from one batch at a time. 
This strategy can be used to force the models focus on a smaller set of techniques, reducing cognitive load and potentially increasing the accuracy of their initial labeling.
After the initial batch-based labeling, each model’s answers were consolidated individually. 
The same model was then re-queried three additional times, but it was restricted to choose only from the techniques it had selected during the initial batch stage. 
This iterative process within each model aimed to refine its previous selections, promoting convergence on the most likely techniques.

\begin{algorithm}
\footnotesize
\caption{Competition questioning}
\label{alg:competition_questioning}

\textbf{Input:} 
\begin{itemize}
    \item \texttt{LLM}: Large Language Model instance
    \item \texttt{NIDS\_Rule}: Network Intrusion Detection System rule to be labeled
    \item \texttt{Techniques}: Complete list of MITRE ATT\&CK techniques
    \item \texttt{Batch\_Size}: Number of techniques per batch
\end{itemize}

\textbf{Output:} 
\begin{itemize}
    \item \texttt{Final\_Labels}: Refined set of techniques associated with the NIDS rule
\end{itemize}

\begin{algorithmic}[1]
\State \textbf{Initialize:}
\State $Selected\_Techniques \gets \emptyset$
\State $Technique\_Batches \gets Divide(\texttt{Techniques}, \texttt{Batch\_Size})$

\State

\State \textbf{Batch-wise Labeling:}
\For{each $Batch \in Technique\_Batches$}
    \State $Predicted\_Techniques \gets Query\_LLM(\texttt{NIDS\_Rule}, Batch)$
    \State $Selected\_Techniques \gets Selected\_Techniques \cup Predicted\_Techniques$
\EndFor

\State

\State \textbf{Iterative Refinement:}
\For{$i \gets 1$ to $R$} \Comment{Number of refinement iterations}
    \State $Refined\_Techniques \gets Query\_LLM(\texttt{NIDS\_Rule}, Selected\_Techniques)$
    \State $Selected\_Techniques \gets Refined\_Techniques$
\EndFor

\State

\State \textbf{Output:}
\Return $Selected\_Techniques$

\end{algorithmic}
\end{algorithm}

Furthermore, \ac{ICL}~\cite{radford2019language,brown2020language} was applied by supplying examples of correctly labeled NIDS rules to demonstrate the desired output format and reasoning process.
A key factor influencing the effectiveness of \ac{ICL} is the quantity of examples provided. 
In general, larger quantity of examples increases the effectiveness of \ac{ICL} jet, the marginal benefit of examples decreases and a saturation can be observed~\cite{liu2021makes}.
Further factors influencing the effectiveness of \ac{ICL} include the distribution of labels within the examples and the ordering of examples~\cite{liu2021makes}.

\subsection{Machine Learning Approach}
\label{sec:ml}
The machine learning approach used the same dataset of Snort rules labeled with MITRE ATT\&CK techniques as the \ac{LLM}-based approach.
However, in the machine learning experiments, both the training and test sets were utilized to build and evaluate the models. 
The dataset was split into train and test sets in a balanced manner to ensure a robust evaluation framework. 
The test set remained the same as the one used for evaluating the \ac{LLMs}, allowing for a direct comparison of performance between the machine learning models and the \ac{LLM}-based approach.

A novel aspect of this study is that the entire machine learning pipeline was designed and executed by the \ac{LLM} itself. 
This process included selecting suitable models, extracting features, writing code, and executing the machine learning tasks. 
The \ac{LLM} had the autonomy to select the best models based on its own knowledge and experience, focusing on maximizing evaluation metrics with an emphasis on the F1-score.

Data preparation involved splitting the dataset into balanced training and test sets to ensure all classes were adequately represented.
For feature extraction, the \ac{LLMs} utilized \ac{TF-IDF} to convert the Snort rules into numerical features. 
The MITRE ATT\&CK technique IDs were binarized into multiple classes, allowing for multi-label classification.
The \ac{LLMs} autonomously identified \ac{TF-IDF} as an appropriate feature extraction method and implemented feature selection techniques to retain the most significant features relevant to the classification task.

The \ac{LLMs} explored various machine learning classifiers suitable for multi-label classification, including \ac{RF}, \ac{SVM}, and \ac{GBM}. 

\begin{algorithm}[h]
\footnotesize
\caption{Machine learning-based approach for Snort rule labeling}
\label{alg:ml_based_approach}
\textbf{Input}
\begin{itemize}
    \item LLM: Large Language Model instance
    \item NIDS\_Rule: Network Intrusion Detection System rule to be labeled
    \item Techniques: Complete list of MITRE ATT\&CK techniques
    \item R: Number of iterations for model tuning
    \item Model\_Type: Set of model types (\ac{RF}, \ac{SVM}, \ac{GBM})
\end{itemize}
    
\textbf{Output}
\begin{itemize}
    \item Final\_Model: Trained machine learning model with selected features
\end{itemize}

\begin{algorithmic}[1]
\State \textbf{Initialize:}
\State Dataset(NIDS\_Rule, Techniques) 
\State Selected\_Features $\gets$ $\emptyset$
\State Model $\gets$ Null
\State 
\State \textbf{Feature Extraction:}
\State TF-IDF\_Features $\gets$ Extract\_TF-IDF(Dataset) 
\State Selected\_Features $\gets$ Feature\_Selection(TF-IDF\_Features) 
\State 
\State \textbf{Model Selection:}
\For{each Model\_Type}
    \State Model $\gets$ Train\_Model(Model\_Type, Selected\_Features, Train\_Set) 
\EndFor
\State 
\State \textbf{Model Tuning:}
\For{$i \gets 1$ \textbf{to} $R$}
    \State Best\_Model $\gets$ Select\_Best\_Model(Models) 
    \State Refined\_Model $\gets$ Refine\_Model(Best\_Model) 
\EndFor
\State 
\State \textbf{Output:}
\Return Best\_Model
\end{algorithmic}
\end{algorithm}



\subsection{Dataset Collection}

To build a dataset comprising labeled \ac{NIDS} rules (e.g. Snort), we collected Snort rules from the Snort community rules repository\footnote{\url{https://www.snort.org/faq/what-are-community-rules}}, focusing on those rules that explicitly referenced MITRE ATT\&CK techniques. 
This involved identifying and extracting rules where their Snort webpage contained a section of MITRE ATT\&CK techniques. 
This approach ensures that the rules contain explicit references to MITRE ATT\&CK techniques, providing a direct link between the detection rules and the adversarial techniques they are designed to identify. 
By selecting rules that included such references, we ensured that the dataset provides a relevant basis for evaluating the performance of various models in the task of automated labeling.

The final dataset comprises \textit{973} rules, each mapped to one or more of \textit{75} unique MITRE ATT\&CK techniques.

\subsection{Experimental Setup}

The dataset was divided into a train and test sets using an 80-20 split, to facilitate the training and evaluation of machine learning models. The training set was used to train the models, while the test set was reserved for the final evaluation to ensure an unbiased performance assessment. 
The distribution of rules across these sets was balanced to ensure a representative sample of different MITRE ATT\&CK techniques in each partition; however, some of the techniques appeared very little in the dataset which made them not feasible to be labeled by the \ac{ML} models (necessitating a split in a test and training set). 
Therefore, we removed rules with less than 5 occurrences prior to the split, which left us with a total of 33 unique techniques across 900 rules - 720 in the train set and 180 in the test. The performance of \ac{LLMs} on the set of 42 rare techniques across 73 rules was also evaluated separately, using the prompt template that achieved the highest F1 score by each \ac{LLM} on the test set.

\begin{figure}[!h]
\includegraphics[width=\textwidth]{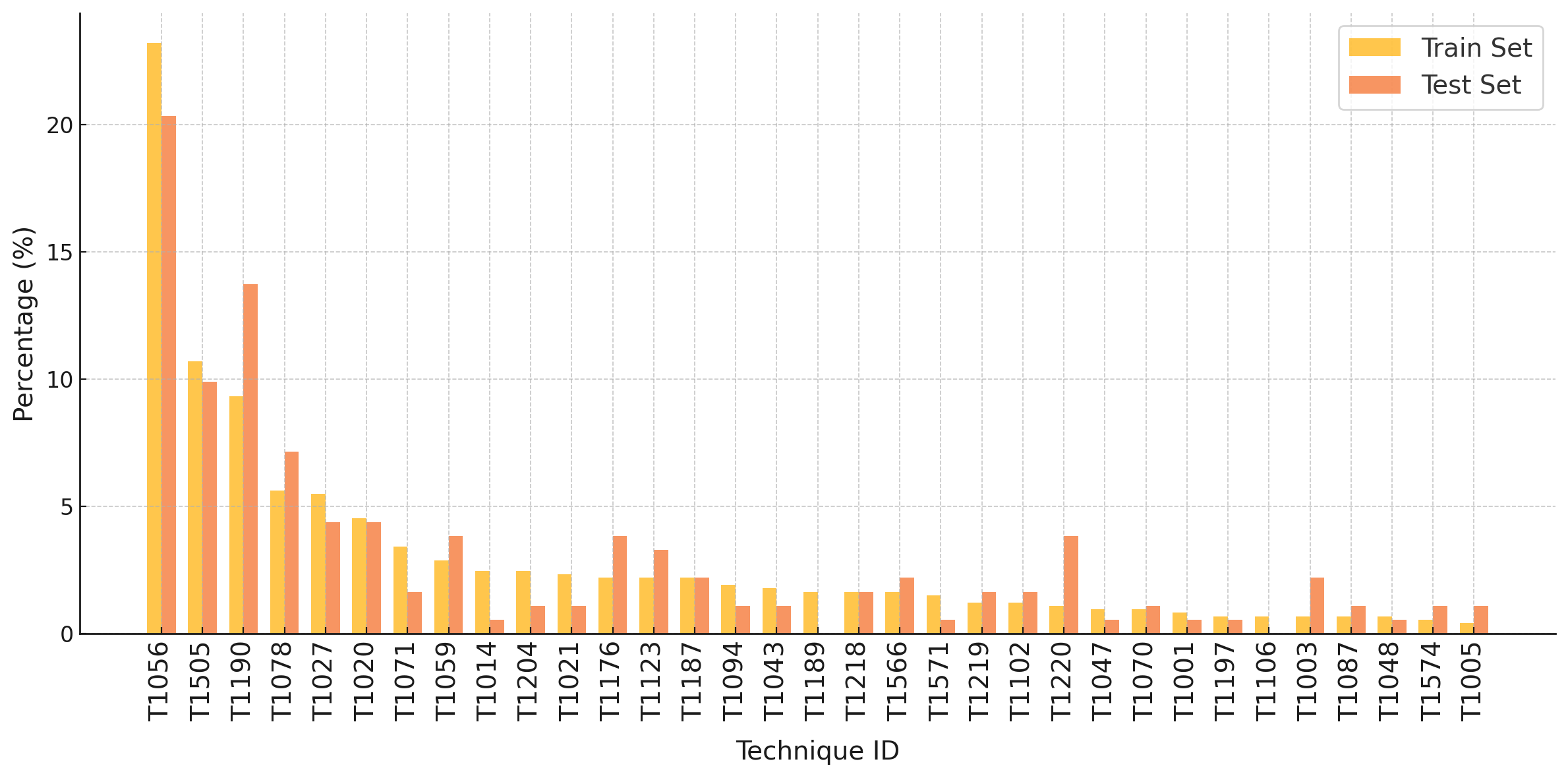}
\caption{Distribution rules across ATT\&CK techniques for both the train and the test sets} \label{fig3}
\end{figure}

\begin{figure}[!h]
\includegraphics[width=\textwidth]{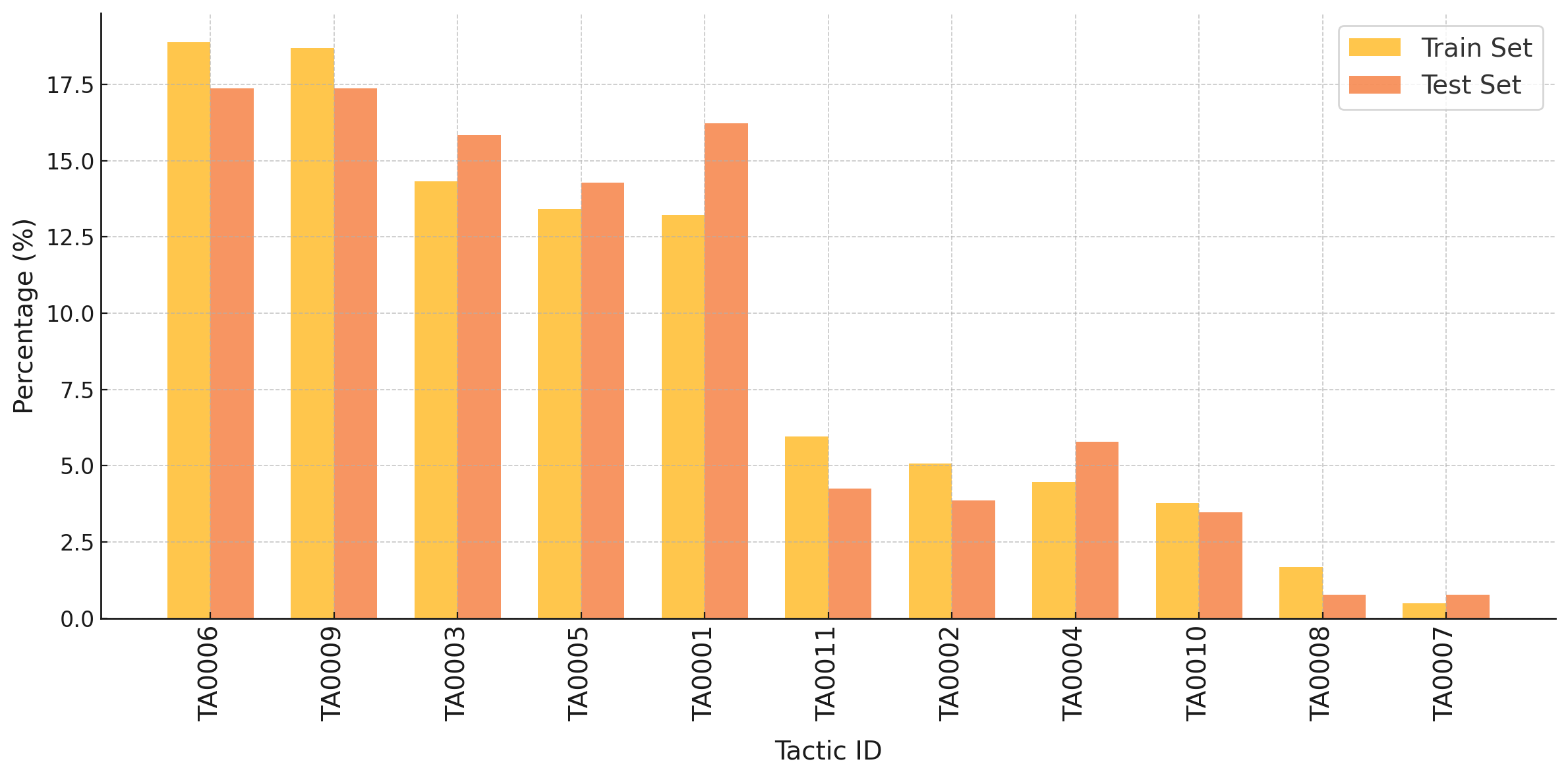}
\caption{Distribution rules across ATT\&CK tactics for both the train and the test sets} \label{fig4}
\end{figure}

Figure~\ref{fig3} presents the percentage of occurrences of each technique out of the total number of occurrences of techniques in the set - for both the train and the test sets. Figure~\ref{fig4} shows the percentage of occurrences of each tactic out of the total number of occurrences of tactics in the set - for both the train and the test sets.

\unskip

\unskip

Three \ac{LLMs} - ChatGPT (chatgpt-4-turbo), Claude (claude-3-sonnet-20240229), and Gemini (gemini-1.0-pro) — were employed to assess their effectiveness in labeling \ac{NIDS} rules with MITRE ATT\&CK techniques.
These models were chosen due to their advanced capabilities in processing complex language tasks, particularly in the context of cybersecurity. 

To evaluate the effectiveness of both the \ac{LLMs} and traditional \ac{ML} approaches in labeling \ac{NIDS} rules, a series of experiments were conducted under varying conditions. 
The experiments were designed to compare the performance of each approach using different configurations and scenarios.

\subsubsection{LLM Experiments}

The experiments conducted encompassed a comprehensive set of scenarios to evaluate the \ac{LLMs}' performance across different conditions. 
These scenarios were combinations of using ($T$) or not using a techniques guide (no techniques, comprehensive set of techniques), providing methodological guidance by few-shot learning (zero - $ICL_0$, one - $ICL_1$, or two - $ICL_2$ examples).
Each combination of these options was tested to thoroughly assess the strengths and weaknesses of each model under varying conditions.

\subsubsection{Machine Learning Experiments}

For the machine learning approach, both the training and test sets were utilized. 
The experiments involved training several classifiers,  and Gradient Boosting Machines, which were selected and configured by the \ac{LLMs} themselves. 
Feature extraction was performed using \ac{TF-IDF}, and hyperparameters tuning was conducted before final testing on the test set.

\subsection{Evaluation Metrics}

The performance of both \ac{LLMs} and machine learning models was assessed using precision, recall, and F1-score. 
These metrics were chosen to provide a balanced evaluation of the models' abilities to correctly label the rules (precision), cover the relevant techniques (recall), and balance both aspects (F1-score). 
Micro-averaging was employed for each evaluation metric described above.

\begin{linenomath}
\begin{equation}
    Precision = \frac{TP}{TP+FP}
\end{equation}
\end{linenomath}

\begin{linenomath}
\begin{equation}
    Recall = \frac{TP}{TP+FN}
\end{equation}
\end{linenomath}

\begin{linenomath}
\begin{equation}
    F1-score = \frac{2 \cdot Precision \cdot Recall}{Precision+Recall}
\end{equation}
\end{linenomath}

We include four baselines which enable to benchmark the results.
Top-1 is defined as a frequency-based baseline.
It describes a heuristic approach of selecting the most frequent technique as a label.
Top-2 consistently describes the heuristic approach of selecting the two top most frequent techniques for each rule as a label.
RT-1 is defined as a random baseline of selecting a random technique of the correct tactic.
We hereby assume that an analyst is able to correctly identify the tactic detected by the \ac{NIDS} rule (as there are models that assist in correctly labeling the \ac{NIDS} rule).
Out of the set of techniques related to the correctly identified tactic, the analyst needs to select a random technique as no method is provided that assists in labeling.
Likewise, RT-2 describes the baseline when selecting two techniques from the set of techniques corresponding to the correct tactic.

\section{Results}
\label{sec:res}
\begin{figure}[!h]
\includegraphics[width=\textwidth]{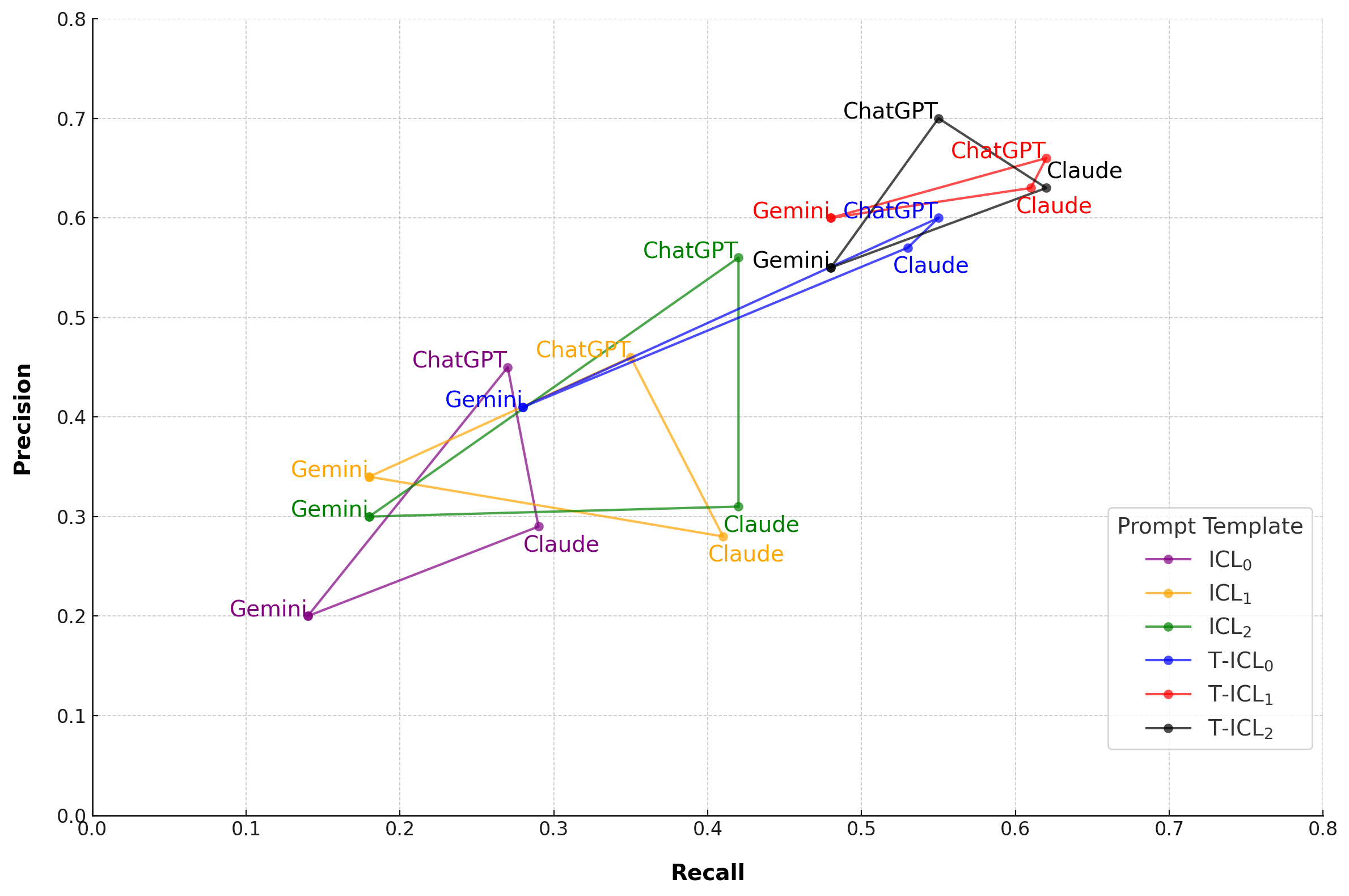}
\caption{Precision and recall of LLM-based technique labeling across different configurations} \label{fig5}
\end{figure}

The performance of both \ac{LLMs} and \ac{ML} models was evaluated in the task of mapping \ac{NIDS} rules to MITRE ATT\&CK techniques furthermore, we conducted a mapping to MITRE ATT\&CK tactics to benchmark with published work. 
Table~\ref{tab:llm-results} and Figure~\ref{fig5} summarize the performance of the tested \ac{LLMs} (Gemini, Claude, and ChatGPT-4) in labeling \ac{NIDS} rules with ATT\&CK tactics and techniques across the different prompt template configurations, while Table~\ref{tab:ml-results} summarizes the precision, recall, and F1-score achieved by the \ac{ML} models trained by the different \ac{LLMs} on the same dataset.
Furthermore, table~\ref{tab:rare-results} provides results of the \ac{LLMs} in labeling rare techniques, where \ac{ML} are not applicable due to a lack of training data.
The performance variations of the chosen baseline heuristics across the test set (see table~\ref{tab:llm-results} and the rare techniques set (see table~\ref{tab:rare-results}) highlight the different underlying complexities of the tasks involved in labeling \ac{NIDS} rules. 
The baselines demonstrate differing effectiveness depending on the structure and characteristics of the set tested on. 
Given this, the test results must be evaluated with these underlying complexities in mind.

\begin{table}[H] 
\caption{Performance of \ac{LLM} models across different configurations.\label{tab:llm-results}}
\newcolumntype{C}{>{\centering\arraybackslash}X}
\begin{tabularx}{\textwidth}{CCCCCCCC}
\toprule
\textbf{Model} & \textbf{Prompt Template} &  & \textbf{Techniques} & & & \textbf{Tactics} &\\ && Precision & Recall & F1-score & Precision & Recall & F1-score\\
\midrule
\hline
\multirow{6}{*}{Gemini} 
 & $ICL_0$ & 0.20 & 0.14 & 0.16 & 0.24 & 0.26 & 0.25 \\
 & $ICL_1$ & 0.34 & 0.24 & 0.28 &0.34 &0.39 &0.36 \\
 & $ICL_2$ & 0.30 & 0.18 & 0.23&0.29 &0.28 &0.29 \\
 & T-$ICL_0$ & 0.41 & 0.28 & 0.33 & 0.38 & 0.45 & 0.41\\ 
 & T-$ICL_1$ & 0.60 & 0.48 & 0.53 & 0.50& 0.66 & 0.57\\ 
 & T-$ICL_2$ & 0.55 & 0.48 & 0.48 & 0.46 & 0.60 & 0.52 \\ 
\hline
\multirow{6}{*}{Claude}
 & $ICL_0$ & 0.29 & 0.40 & 0.34&0.39 &0.57 &0.46 \\
 & $ICL_1$ & 0.28 & 0.41 & 0.33&0.39 &0.60 &0.47 \\
 & $ICL_2$ & 0.31 & 0.42 & 0.36& 0.42 & 0.64 & 0.51 \\
 & T-$ICL_0$ & 0.51 & 0.59 & 0.55 & 0.57 & 0.79 & 0.66\\ 
 & T-$ICL_1$ & 0.57 & 0.60 & 0.59 & 0.59 & \textbf{0.80} & 0.68\\ 
 & T-$ICL_2$ & 0.63 & \textbf{0.61} & \textbf{0.62} & 0.61 & \textbf{0.80} & \textbf{0.69}\\ 
\hline
\multirow{6}{*}{ChatGPT}
 & $ICL_0$ & 0.45 & 0.27 & 0.34 & 0.50 & 0.45 & 0.47 \\
 & $ICL_1$ & 0.46 & 0.29 & 0.36 & 0.49 & 0.47 & 0.48 \\
 & $ICL_2$ & 0.35 & 0.30 & 0.32 & 0.44 & 0.50 & 0.47\\ 
 & T-$ICL_0$ & 0.56 & 0.55 & 0.56 & 0.62 & 0.69 & 0.65\\ 
 & T-$ICL_1$ & 0.66 & 0.58 & \textbf{0.62} & 0.62 & 0.69 & 0.65\\ 
 & T-$ICL_2$ & \textbf{0.70} & 0.55 & \textbf{0.62} & \textbf{0.66} & 0.72 & \textbf{0.69}\\ 
\hline
\\
& \textbf{Approach}	& 	& &  &  &  & \\
\hline
\multirow{4}{*}{\textbf{Baseline}} & Top-1	& 0.2	& 0.2	& 0.2 & 0.25 & 0.34 & 0.29 \\
& Top-2	& 0.17	& 0.34	& 0.22 & 0.24 & 0.5 & 0.33 \\
& RT-1	& 0.06	& 0.05	& 0.05 & N/A & N/A & N/A\\
& RT-2	& 0.12	& 0.1	& 0.1 & N/A & N/A & N/A\\
\hline
\bottomrule
\end{tabularx}
\end{table}

The presented methodologies significantly outperform the baselines across all prompt template configurations besides Gemini $ICL_0$. 
This shows the general feasibility of applying \ac{LLMs} within the task of labeling \ac{NIDS} rules with MITRE ATT\&CK techniques and tactics.

The results demonstrate a clear hierarchy in model performance, with Claude and ChatGPT consistently outperforming Gemini across all tested configurations.
Claude and ChatGPT exhibited superior performance, achieving higher precision and recall in their responses, regardless of the prompt template used.
The highest F1-score for technique (0.62) and tactic labeling (0.69) were reached under the T-$ICL_2$ configuration, which provided example-driven contextualization and methodological guidance.
While ChatGPT provides superior results with regard to precision, Claude has its strength in providing higher levels of recall.

Unlike Claude and ChatGPT, Gemini demonstrated the highest performance within the T-$ICL_1$ setup (with an F1 score of 0.53 for technique labeling and 0.57 for tactics labeling).
Furthermore, Gemini showed the highest performance increase when combining contextual and methodological guidance. 

Within our experiments we reached significantly higher performance than previous works reaching maximal F1-score of 0.32 using the same model of ChatGPT employed in this work as well as with ChatGPT-3.5 where an F1-score of 0.49 has been reached~\cite{daniel2023labeling}.
While Daniel et al.~\cite{daniel2023labeling} combine labels of different results our workflow allows to reach more precise results through improved prompt design.

\unskip

When comparing the different configurations of the prompt template, the contextual guidance shows a greater effect on the probability that the \ac{LLM} selects the right label (the T-$ICL_0$ prompt template outperforms $ICL_2$).
The superior performance with the T-$ICL_2$ prompt template demonstrates the value of providing both contextual information and methodological guidance (specifically \ac{ICL}) within the prompts.

Table~\ref{tab:rare-results} shows the results attained by the \ac{LLMs} (with their best performing prompt template configuration as demonstrated using the test set; see table~\ref{tab:llm-results}) when labeling is performed on the rare techniques set.
Gemini, Claude and ChatGPT show comparable performance within their best performing configurations.
Overall the reached performance of the \ac{LLMs} in labeling rare techniques is low. 
However, compared to the baselines computed, the \ac{LLMs} perform superior proofing their value for labeling rarely seen techniques (e.g. if labeled data is scarce).
Gemini performs especially good in labeling tactics, suggesting that closely related techniques (techniques of the same tactic) were frequently proposed in technique labeling task.

\begin{table}[H] 
\caption{Performance of LLMs in rare technique labeling.\label{tab:rare-results}}
\newcolumntype{C}{>{\centering\arraybackslash}X}
\begin{tabularx}{\textwidth}{CCCCCCCC}
\toprule
\textbf{Model} & \textbf{Prompt Template} & & \textbf{Techniques} & & & \textbf{Tactics} &\\ &&Precision & Recall & F1-score& Precision & Recall & F1-score\\
\midrule
\hline
Gemini		& T-$ICL_1$	& 0.22	& \textbf{0.21}	& 0.22 & \textbf{0.29}	& \textbf{0.38}	& \textbf{0.33}\\
Claude		& T-$ICL_2$	& 0.23	& 0.18	& 0.20 & 0.26 & 0.29 & 0.27\\
ChatGPT		& T-$ICL_2$	& \textbf{0.29}	& 0.19	& \textbf{0.23} & 0.28 & 0.26 & 0.27\\
\hline
\\
& \textbf{Approach}	& 	& &  &  &  & \\
\hline
\multirow{4}{*}{\textbf{Baseline}} & Top-1	& 0.05	& 0.05	& 0.05 & 0.09 & 0.09 & 0.09 \\
& Top-2	& 0.05	& 0.11	& 0.08 & 0.13 & 0.22 & 0.17 \\
& RT-1	& 0.04	& 0.04	& 0.04 & N/A & N/A & N/A\\
& RT-2	& 0.09	& 0.09	& 0.09 & N/A & N/A & N/A\\
\hline
\bottomrule
\end{tabularx}
\end{table}

\begin{table}[H] 
\caption{Performance of ML models developed by the different LLMs.\label{tab:ml-results}}
\newcolumntype{C}{>{\centering\arraybackslash}X}
\begin{tabularx}{\textwidth}{CCCCCCCC}
\toprule
\textbf{Developing Model} & \textbf{ML Model} & & \textbf{Techniques} & & & \textbf{Tactics} &\\ &&Precision & Recall & F1-score& Precision & Recall & F1-score\\
\midrule
\hline
Gemini		& SVM			& 0.88			& \textbf{0.87}			& \textbf{0.87} & 0.91 & \textbf{0.92} & \textbf{0.92}\\
Claude		& SVM			& 0.86			& 0.85			& 0.85 & 0.91 & \textbf{0.92} & 0.91\\
ChatGPT		& SVM			& \textbf{0.92}			& 0.70			& 0.79& \textbf{0.96} & 0.77 & 0.85\\
\hline
\bottomrule
\end{tabularx}
\end{table}

Within our tests, all \ac{ML} models outperformed the \ac{LLMs} across all metrics. 
The \ac{SVM} model trained by Gemini on the labeled dataset achieved the highest overall performance in technique labeling, with a precision of 0.88, recall of 0.87, and an F1-score of 0.87. 
The \ac{SVM} model trained by Claude achieved an F1-score of 0.85, supported by a precision of 0.86 and recall of 0.85. 
Finally, \ac{SVM} trained by ChatGPT performed worse than the other models, achieving an F1-score of 0.79, with the highest precision of 0.92 and the lowest recall of 0.70.
For tactic labeling, the models trained by Claude (F1-score of 0.91) and Gemini (F1-score of 0.92) showed competitive results.
Slightly worse the \ac{SVM} trained by ChatGPT performs at a F1-score of 0.85.

Overall, the \ac{ML} models exhibited greater consistency and robustness compared to the \ac{LLMs}.
For labeling tactics, our approach achieves a F1-score of 0.92 (ML-model trained and developed by Gemini), demonstrating comparable performance to previously published tools in labeling attack tactics~\cite{lin2022attack}.
The results indicate strong precision, recall, and F1 scores for labeling \ac{NIDS} rules with tactics, providing comparable performance to existing \ac{ML}-based methods for labeling \ac{NIDS} rules with tactics~\cite{lin2022attack}.
Furthermore, the presented models provide a good basis for decreasing the alert explainability gap by providing highly precise tactic labels for \ac{NIDS} rules.

\section{Discussion}
\label{sec:disc}

The results highlight clear differences between the \ac{LLM}-based and \ac{ML}-based approaches for mapping \ac{NIDS} rules to MITRE ATT\&CK, both in terms of performance metrics and potential practical applications.
We reached comparable results as presented by related work with regard to tactic labeling~\cite{lin2022attack} while designing the workflow for technique labeling.
Furthermore, we introduce the first work elaborating on the labeling of \ac{NIDS} rules with MITRE ATT\&Ck techniques and contribute a methodology for labeling based on \ac{LLMs} and \ac{ML}-models.

\subsection{LLM Performance and Insights}

The \ac{LLMs} demonstrated variable performance depending on the configuration and prompt design.
Highlighting the need for efficient prompt engineering. 

The results of our comparative evaluation of Claude, ChatGPT, and Gemini with regard to their capabilities in the context of \ac{NIDS} rule labeling across multiple prompt configurations reveal significant performance discrepancies. 
Claude and ChatGPT consistently outperformed Gemini in terms of recall and precision. 
Gemini exhibited the poorest performance in all configurations tested.
These findings warrant further exploration of the factors influencing the observed performance variations.
Furthermore, in future research additional prompt design techniques and \ac{LLMs} should be tested to further boost performance of \ac{LLMs} for the proposed labeling task allowing the practical application within \ac{SOCs}.

The reached performances of Claude and ChatGPT reaching F1 scores greater than 0.6 underscores the utility of incorporating using \ac{LLMs} for labeling techniques. 
Furthermore, with regard to the prompt templates the superiority of the T-$ICL_2$ templates proves the benefits of example-driven prompts that guide the model toward task-specific reasoning.
Yet, while \ac{ICL} allows \ac{LLMs} to leverage examples provided directly within the prompt to improve performance on a given task, it may not always be the most efficient strategy, particularly in resource-constrained environments. 
One key limitation of \ac{ICL} is that it increases computational costs and token usage, especially when large examples or numerous prompt tokens are required. 
This results in higher compute costs and longer response times, which can be prohibitive in large-scale or real-time applications.

\subsection{ML Model Superiority}

The \ac{ML} models demonstrated superior performance across all datasets achieving the highest F1-scores (0.87 for technique labeling and 0.92 for tactic labeling) when trained by Gemini. 
This result emphasizes the strength of supervised learning, particularly when trained on well-annotated datasets. 
The \ac{ML} models’ precision scores, such as 0.92 for the \ac{SVM} model trained on ChatGPT in technique labeling and 0.96 in tactic labeling, were notably higher than those of the \ac{LLMs}, reflecting their ability to minimize false positives effectively.

The strong performance of \ac{ML} models can be attributed to their capacity to exploit structured feature representations, such as \ac{TF-IDF} vectors, to identify patterns within the dataset. 
However, it is worth noting that the success of the \ac{ML} approach heavily depends on the quality of the training data. 

\subsection{Comparative Implications}

While \ac{LLMs} offer flexibility and the ability to generate explanations for their predictions, their lower performance metrics suggest that they are at least currently better suited for tasks where interpretability and generalization are more important than raw accuracy. 
On the other hand, \ac{ML} models excel in precision and recall but rely on high-quality labeled datasets and cannot inherently provide reasoning for their decisions.
Yet, if no labeled datasets are available \ac{ML} models can be trained on, \ac{LLMs} can provide significant value in early stages of generating labeled datasets especially if the \ac{LLMs} are used to assist human analysts by generating candidate labels (human in the loop processes).
LLMs can thereby take advantage of their ability to understand \ac{NIDS} rules and generate labels especially in setups with partial information.

A hybrid approach could harness the strengths of both methods: \ac{LLMs} could be employed for initial labeling and to provide contextual explanations, while \ac{ML} models could refine these labels for deployment in high-stakes applications requiring high precision and recall. 
Future work could explore ensemble strategies or the integration of \ac{LLM}-generated insights into \ac{ML} feature engineering to further enhance performance.


\section{Conclusions}
\label{sec:conc}

Within this study we present a dataset comprising 973 labeled \ac{NIDS} rules.
The collected dataset serves as a useful resource for the cybersecurity research community, providing a ground truth for developing and testing new methods for mapping \ac{NIDS} rules to MITRE ATT\&CK techniques.
This dataset can be utilized in future studies to benchmark model performance and explore new approaches to enhancing \ac{NIDS} capabilities.

The experiments we conduct on the basis of this dataset underscore the potential of leveraging both \ac{LLMs} and \ac{ML} techniques to automate the labeling of \ac{NIDS} rules with MITRE ATT\&CK techniques and tactics. 
LLMs such as ChatGPT, Claude, and Gemini demonstrated their ability to provide contextual and explainable mappings, particularly when optimized through prompt engineering and contextual guidance. 
Among these, ChatGPT achieved the best overall performance, highlighting its potential to assist security analysts in tasks requiring contextual reasoning. 
However, their precision and recall scores, while promising, were consistently outperformed by supervised \ac{ML} models trained on the same dataset.

The \ac{ML} models, particularly \ac{SVM}, showcased superior accuracy, benefiting from well-defined feature extraction and robust training datasets generated by the \ac{LLMs}. 
The high precision and recall achieved by the \ac{ML} models affirm their applicability for high-stakes scenarios where accuracy is critical. 
However, these models rely heavily on high-quality, pre-labeled datasets, a limitation that could restrict scalability to broader domains without significant initial manual effort.

A key contribution of this study is the collection of a large dataset of 973 Snort \ac{NIDS} rules mapped to MITRE ATT\&CK techniques and tactics. 
This dataset not only serves as a foundation for the evaluations conducted in this work but also provides a valuable resource for the cyber security research community. 
It can be leveraged for training, benchmarking, and further improving both \ac{LLM}- and \ac{ML}-based methods for \ac{CTI} enrichment.

A key finding of this study is the complementary nature of \ac{LLMs} and \ac{ML} models. 
While \ac{LLMs} provide a flexible and scalable framework for generating initial labels and explanations, \ac{ML} models excel in refining these labels to achieve superior performance metrics. 
This synergy suggests the feasibility of a hybrid approach where \ac{LLMs} are used for initial data enrichment and add explainability and \ac{ML} models for final labeling.

Building on these findings, several directions can be pursued:

\begin{itemize}
    \item \textbf{Prompt engineering:} Further prompt engineering techniques could be used to increase performance. 
    Furthermore, the effect of improving contextual guidance by restricting the technique set to relevant techniques detectable with \ac{NIDS} (e.g. selected based on DETT\&CT) should be investigated.  
    \item \textbf{Hybrid Approaches:} Future research could explore hybrid frameworks that integrate \ac{LLMs} for initial rule labeling and \ac{ML} models for fine-tuning. 
    This approach may combine the scalability of \ac{LLMs} with the precision of \ac{ML} models.
    
    \item \textbf{Domain-Specific Fine-Tuning:} Fine-tuning \ac{LLMs} with domain-specific datasets could improve their accuracy and reduce the need for extensive prompt engineering. 
    This would be particularly beneficial for complex domains such as \ac{ICS}.
    
    \item \textbf{Enhanced Feature Engineering:} \ac{ML} models could benefit from incorporating additional features, such as contextual relationships within rules or temporal correlations, to further boost their labeling accuracy.
    
    \item \textbf{Evaluation on Diverse Datasets:} Expanding the evaluation to include datasets from different domains and attack scenarios will ensure generalizability and robustness of the proposed methods.
    
    \item \textbf{Explainable AI:} While \ac{LLMs} inherently provide explainability through natural language outputs, enhancing \ac{ML} models with interpretable explanations could improve trust and adoption in operational environments.
\end{itemize}

In conclusion, this study demonstrates that automated labeling of \ac{NIDS} rules is both feasible and effective using a combination of \ac{LLMs} and \ac{ML} techniques. 
By addressing the challenges of scalability, accuracy, and domain adaptation, the proposed approaches can significantly enhance the capabilities of cyber security analysts in mitigating evolving threats.

\section{Abbreviations}{
The following abbreviations are used in this manuscript:
\small
\begin{acronym}[MCML-C] 
\acro{AT}{Attack Type}
\acro{AI}{Artificial Intelligence}
\acro{CTI}{Cyber Threat Intelligence}
\acro{DT}{Decision Tree}
\acro{GBM}{Gradient Boosting Machine}
\acro{GPTs}{Generative Pre-trained Transformers}
\acro{HIDS}{Host-based Intrusion Detection Systems}
\acro{ICL}{In-Context-Learning}
\acro{ICS}{Industrial Control Systems}
\acro{IDS}{Intrusion Detection System}
\acro{IoC}{Indicators of Compromise}
\acro{E}{Explanation}
\acro{EDR}{Endpoint Detection and Response}
\acro{KNN}{K-Nearest Neighbor}
\acro{LCA}{Lowest Common Ancestor}
\acro{LLM}{Large Language Model}
\acro{LLMs}{Large Language Models}
\acro{MCML-C}{Multi-Class Multi- Label Classifier}
\acro{ML}{Machine Learning}
\acro{ML-C}{Multi Label Classification}
\acro{NER}{Named Entity Recognition}
\acro{NIDS}{Network Intrusion Detection Systems}
\acro{RF}{Random Forest}
\acro{SOCs}{Security Operations Centers}
\acro{SVM}{Support Vector Machines}
\acro{T}{Technique}
\acro{TA}{Tactic}
\acro{TH}{Threat Hunting}
\acro{TF}{Term Frequence}
\acro{TF-IDF}{Term Frequence Inverse Document Frequency}
\acro{TTPs}{Tactics, Techniques, and Procedures}
\end{acronym}
}

\bibliographystyle{unsrt}  
\bibliography{references}  






\end{document}